\documentclass[marginparwidth=30pt,a4paper,12pt]{article}

\usepackage[utf8]{inputenc}
\usepackage{graphicx}
\usepackage{amsmath}
\usepackage{gensymb}
\usepackage{multirow}
\usepackage{cleveref}
\usepackage{caption}
\usepackage{float}
\usepackage{xcolor}
\usepackage{authblk}
\usepackage[style=nature,doi=false,isbn=false,url=false,eprint=false]{biblatex}
\usepackage{comment}

\title{Scaling, saturation, and upper bounds in the failure of topologically interlocked structures}
\author[1]{Shai Feldfogel}
\author[2]{Konstantinos Karapiperis}
\author[3]{Jose Andrade}
\author[1]{David S. Kammer}
\affil[1]{Institute for Building Materials, ETH Zurich, Switzerland}
\affil[2]{Department of Mechanical and Process Engineering, ETH Zurich, Switzerland}
\affil[3]{Department of Mechanical and Civil Engineering, Caltech, Pasadena, California, USA}

\begin{document}

\begin{comment}

Why (purpose)

Who (audience)

What (Main messages)

\end{comment}

\maketitle

\begin{abstract}
Topological Interlocking Structures (TIS) have been increasingly studied in the past two decades.
However, some fundamental questions concerning the effects of Young's modulus and the friction coefficient on the structural mechanics of the most common type of TIS application - centrally loaded slabs - are not yet clear. 
Here, we present a first-of-its-kind parametric study based on the Level-Set-Discrete-element-Method that aims to clarify how these two parameters affect multiple aspects of the behavior and failure of centrally-loaded TIS slabs.
This includes the evolution of the structural response up to and including failure, the foremost structural response parameters, and the residual carrying capacity.
We find that the structural response parameters in TIS slabs scale linearly with Young's modulus, that they saturate with the friction coefficient, and that the saturated response provides an upper-bound on the capacity of centrally loaded TIS slabs reported in the literature.
This, together with additional findings, insights, and observations, comprise a novel contribution to our understanding of the interlocked structural form.
\end{abstract}

\section{Introduction} \label{sec:Introduction}
\begin{figure} 
    \includegraphics[width=1\textwidth]{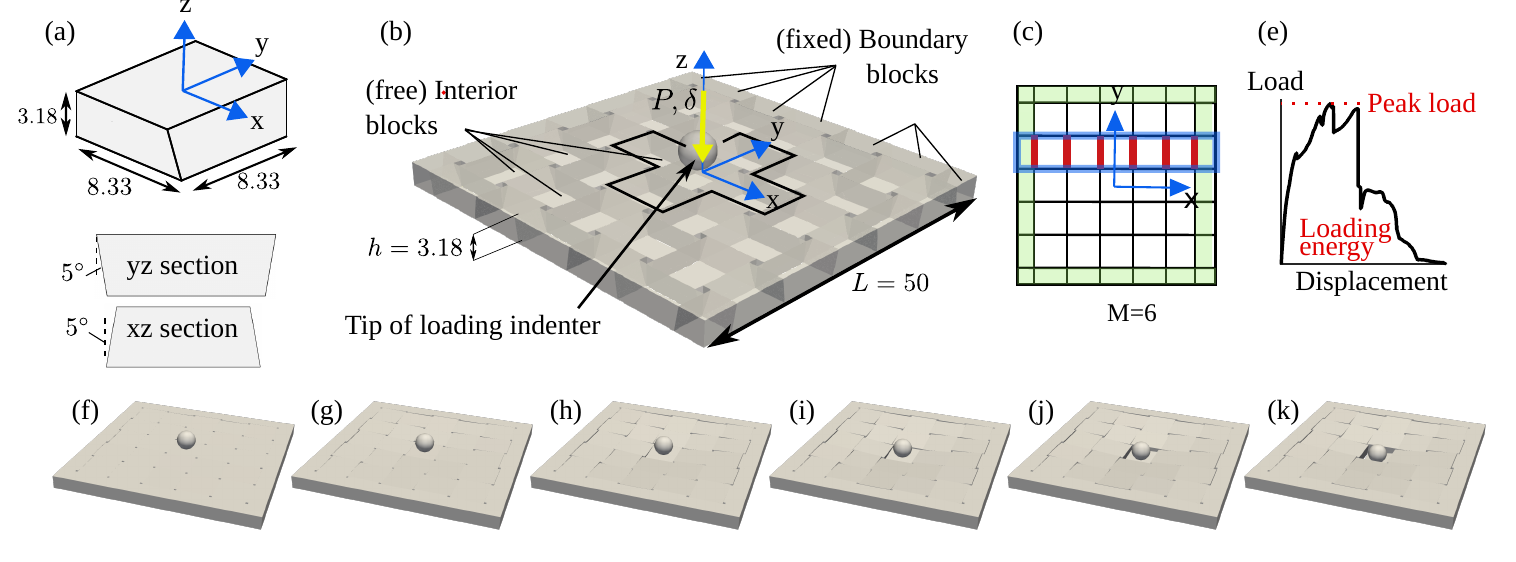} 
    \centering
    \caption{Examined configuration - (a) a typical internal block and cross-sections in the xz and yz planes; (b) examined slab configuration with boundary and loading conditions. The force P exerted by the indenter on the slab and the corresponding indenter displacement $\delta$ are indicated by a yellow arrow in the -z direction; (c) a typical strip showing that $M=6$; (d) a typical load-deflection curve indicating in red the peak load and the loading energy; and (f-k) sequential snapshots illustrating the typical response of TIS slabs to centrally applied loads.}
    \label{fig:configuration}
\end{figure}

Topological Interlocking Structures (TIS) are assemblies of un-bonded building blocks that obtain their structural integrity through the blocks' interlocking geometries and through the contact and friction interactions that develop at their interfaces.
The most common and widely studied TIS application is TIS slabs, under quasi-statically-applied indentation loads \cite{dyskin_fracture_2003,dyskin_topological_2003,dyskin_principle_2005,molotnikov_percolation_2007, carlesso_enhancement_2012, carlesso_improvement_2013,dyskin_topological_2019,dyskin_mortarless_2012, estrin_design_2021,estrin_architecturing_2021}.
Fig \ref{fig:configuration}(b) illustrates a centrally loaded TIS slab.
Fig \ref{fig:configuration}(f-k) illustrates through sequential snapshots the typical response of such slabs as observed in experiments and Fig. \ref{fig:configuration}(e) depicts a typical load-deflection curve.

As seen in Fig. \ref{fig:configuration}(f-h), under small loads, the blocks largely stick against one another and the slab deforms  as a whole. 
We refer to this type of structural action as stick-governed \cite{schaare_point_2008,autruffe_indentation_2007}.
As the load increases, slipping mechanisms initiate near the loaded block and gradually become more noticeable, see Fig. \ref{fig:configuration}(i,j).
Eventually, block/s start to fall-off from the assembly, and the load drops to zero which mark the structural failure, see Fig. \ref{fig:configuration}(k).
We refer to this type of structural action as slip-governed.
The stick-governed initial response and the slip-governed failure depicted in Fig. \ref{fig:configuration}(f-k), is, by far, the most prevalent one observed in experiments \cite{dyskin_fracture_2003,mirkhalaf_toughness_2019,khandelwal_transverse_2012,krause_mechanical_2012,feng_impact_2015,molotnikov_percolation_2007,djumas_deformation_2017}.
Therefore, sound understanding of the stick governed regime, and most importantly, of the slip-governed one which precedes the fall-off of blocks is key to the study and design of TIS slabs.

Since TIS slabs transmit loads through contact and friction forces between the blocks, the main material parameters that govern the interplay between the stick- and slip-regimes, and thus the behavior and failure of TIS slabs as a whole, are Young's modulus $E$ and the interfaces' ability to resist sliding \cite{dugue_indentation_2013,feng_impact_2015,khandelwal_scaling_2013}.
$E$ is important mainly because it governs the slab's in-plane compressive stiffness which in turn governs the in-plane forces, the global stiffness, and the peak load.
The interfaces' ability to resist sliding, commonly represented by the friction coefficient $\mu$, is important because it determines the transition to and the evolution of the slip-mechanisms that eventually lead to blocks' fall-off and structural failure.

The importance of $E$ and $\mu$ in governing the structural action of TIS slabs raises fundamental questions that are at focus of the present study: How does the global structural response parameters, namely the peak load, the loading energy, and the initial stiffness scale with $E$ and $\mu$?
How do $E$ and $\mu$ affect the extent of slip-mechanism-induced damage in terms of the number of blocks that fall-off and the commensurate residual load carrying capacity?
Is there a saturation level of $E$ and $\mu$ beyond which they do not further contribute to the carrying capacity or affect the nature of the failure mechanism? 
Is there an upper bound to the carrying capacity, and if so, how is related to $E$ and $\mu$?

As we show next, these fundamental questions have not yet been systematically addressed in the literature\footnote{We attribute this somewhat surprising fact to the difficulty of capturing and quantifying the slip-governed response of TIS using the Finite Element Method (FEM) -- the most widely-used structural analysis tool.}.
The main objective of this paper is to address these open questions and thereby to provide new fundamental insights into the behavior and failure of TIS slabs.

Heretofore, the majority of the studies that considered the effects of $E$ and $\mu$ on the behavior of TIS slabs focused on the stick-governed regime. Most of these studies employed the Finite Element Method (FEM) \cite{autruffe_indentation_2007,schaare_point_2008,feng_impact_2015,dugue_indentation_2013}, others used an analytical model \cite{khandelwal_scaling_2013} based on the thrust-line analogy \cite{khandelwal_transverse_2012,khandelwal_scaling_2013,short_scaling_2019}, and yet another took a direct experimental approach \cite{schaare_point_2008}. 
These studies generally concur that the initial global stiffness and the peak load in the stick-governed regime increase approximately linearly as a function of $E$.
Concerning the effects of $\mu$, it was found experimentally that the peak load \cite{schaare_point_2008} and the initial global stiffness increased while the loading energy decreased as a function of $\mu$ \cite{autruffe_indentation_2007}.

By considering a wider range of $\mu$'s than in the aforementioned experimental studies, Khandelwal et al. \cite{khandelwal_scaling_2013} found, using FEM analyses, that the initial global stiffness grows as a function of $\mu$ until, beyond a certain value of $\mu$, it becomes saturated, meaning it does not further grow.
Dugue et al. \cite{dugue_indentation_2013} observed a similar stiffness saturation.
In both these studies, the phenomenon of saturation beyond a certain $\mu$ was not explored in terms of the peak load and the loading energy, which are arguably more significant response parameters, or in terms of the type of failure mechanism.
While the studies that focused on the stick-governed regime provide data on how $E$ and $\mu$ affect the response parameters on TIS, they do not address the slip-governed regime which is the one that features much more prominently in TIS failure.
Even within the stick-goverened regime, the treatment is partial in that it only considered a relatively small ranges of $E$ and $\mu$ and the latter's effects on only few of the response parameters of interest.

A couple of studies considered the effects of $E$ and $\mu$ in more realistic context of TIS' slip-governed failure.
Feng et al. \cite{feng_impact_2015} studied the slip-governed response of a TIS slab to a central impact load.
They found based on FEM analyses that all the aforementioned response parameters generally grow with $\mu$'s in the range 0.1-1.
However, more quantitative information on each of these parameters scales with $\mu$, as well as the effect of $E$ was not discussed.
Also, the more commonly occurring scenario of quasi-static loading, which is likely to have a different dependence on $E$ and $\mu$, was not considered.

Koureas et al. \cite{koureas_failure_2022} studied the effects $E$ and $\mu$ in the more commonly-considered case of quasi-static loading in the context of a TIS-like beam configuration.
This study showed that the peak load, loading energy, and initial global stiffness scale linearly with $E$ and increase sub-linearly as a function of $\mu$ until they saturate for large $\mu$'s.
It also showed that this saturated structural response corresponds to a failure mechanism that is stick-governed.
While providing new observations and interesting clues relating to the open question that are at the focus of the present study, the limitation of \cite{koureas_failure_2022} in the context of the present study is that it did not consider TIS slabs but a simpler beam configuration.
The more complex two-way structural action of TIS slabs does not allow to apply the beam-based results in \cite{koureas_failure_2022} to TIS slabs.

\begin{comment}
One example demonstrating the qualitative differences between TIS beams and TIS slabs for which a separate treatment is required for the latter is that TIS slabs can develop larger normalized deflections and commensurate loading energy. Specifically, the maximal deflection in TIS beams is always smaller than the height of the beam \cite{dalaq_manipulating_2020,dalaq_strength_2019,koureas_failure_2022}, whereas in TIS slabs it often exceeds the slab's thickness \cite{krause_mechanical_2012,feng_impact_2015, mirkhalaf_toughness_2019}, sometimes by a factor larger than two \cite{djumas_deformation_2017}.    
\end{comment}

In summary, while some of the effects of $E$ and $\mu$ on the behavior of TIS slabs have been sporadically addressed in the literature, the existing studies focused on the pre-failure stick-governed regime, they considered a relatively small range of $E$ and $\mu$, and they only looked at one of the structural response parameters at a time.
The one study that more fully addressed the effects of $E$ and $\mu$ within the slip-governed regime was limited to TIS-like beams and did not address the richer two-way action of TIS slabs, which are the more common TIS application.
The main objectives of this study are to help bridge this gap and to shed new light on the effects of $E$ and $\mu$ on the behavior and failure of TIS slabs.
To obtain these objective, we a perform a numerical parametric study using the Level-Set-Discrete-Element-Method (LS-DEM). 
A numerical approach allows to isolate the effects of $E$ and $\mu$, as well as to explore a wide and independent range of these parameters.

\section{Methodology - an LSDEM approach} \label{sec:Methodology}
LSDEM, originally developed for granular mechanics applications  \cite{kawamoto_level_2016}, has recently been adapted for structural analysis of TIS \cite{feldfogel_discretization-convergent_2022}, and validated against experimental results of TIS slabs \cite{feldfogel_failure_2022}.
It's main advantage compared with FEM is that it is well-equipped to capture the slip-governed response of TIS while closely estimating the load-deflection curve.
The concepts underlying the application of LS-DEM to structural analysis of TIS have been detailed elsewhere \cite{feldfogel_discretization-convergent_2022}, so, we will mention here only the most essential features.

LS-DEM's application to TIS involves a seeming contradiction.
On the one hand, TIS are (by definition) infinitely strong and stiff when the blocks' $E$ is infinite.
On the other hand, our LS-DEM model operates under the assumption of non-breakable rigid-body blocks.
Therefore, on its face, modeling TIS with LS-DEM should always yield infinite global stiffness and carrying capacity.
The fortunate reason that this is not the case is that the contact between blocks is enforced in LS-DEM in a penalty sense, which invariably involves small penetrations between blocks.
The cumulative result of these penetrations is effective global compressibility and commensurate finite global stiffness and carrying capacity \cite{feldfogel_failure_2022}.

The main concept underlying our model is to harness this built-in mechanism of effective global compressibility and to relate it to the true elastic compressibility of a given TIS slab. 
Accordingly, we consider the penetration penalty parameter in our model $k_n$ as a proxy of the true $E$ and we calculate it as:
\begin{equation} \label{eq:k_n}
    k_n=E \cdot \frac{M}{L}
\end{equation}
where $M$ is the number of interfaces in a single row of blocks and $L$ is the side length of the considered slab.
Using a $k_n$ determined from Eq. (\ref{eq:k_n}), it has been shown in \cite{feldfogel_failure_2022} that the LS-DEM model has the same effective in-plane stiffness as a TIS slab with an elastic modulus $E$.
In what follows, we shall adhere to Eq. (\ref{eq:k_n}) so that $E$ and $k_n$ are always directly relatable by the constant factor $M/L$, which is always known for a given slab.
Further details concerning our LS-DEM model for TIS -- additional assumptions, the rational behind it, and its ability to capture the experimentally observed behavior and failure of TIS -- are available in \cite{feldfogel_failure_2022}.

\section{Results} \label{sec:Results}
\subsection{Examined configuration and numerical modeling} \label{subsec:Examined configuration and numerical modeling}
The examined configuration follows the experimental set-up in \cite{mirkhalaf_toughness_2019} and it is depicted in Figure \ref{fig:configuration}(a,b).
In the slab's plan view shown in Fig. \ref{fig:configuration}(c), red line segments indicate the six interfaces in a single row of blocks, so that $M/L=6/50$ [1/mm].
An illustrative $P-\delta$ curve indicating the main response parameters we examine is shown in Fig. \ref{fig:configuration}(d).

We chose the set-up described in Figure \ref{fig:configuration}(a,b) for the $E$-$\mu$ parametric investigation for two reasons. 
First, square slabs under central indentation loading is the most common set-up studied in the TIS literature.
Second, among many possible block geometries, planar-faced blocks with an angle of inclination of 5$^{\circ}$ have been shown to lead to the highest carrying capacities among a variety of different block geometries \cite{koureas_failure_2022}.
In that sense, they provide an upper-bound on the block-geometry degree of freedom.
Additional details on the configuration and on the LS-DEM simulations are available in \cite{feldfogel_failure_2022,feldfogel_discretization-convergent_2022}.

Regarding the numerical modeling, the slab was quasi-statically loaded by prescribing a constant velocity to a 2.5 mm spherical indenter.
A rate of 6 mm/sec was found to be low enough so as to not introduce any observable dynamical effects.
For our analyses, we first applied gravity load until the assembly reached a relaxed state, which we then took as the initial configuration for the indentation phase.
The material density and the friction coefficient were taken equal to 2500 kg/m$^3$ and $\mu$=0.23, respectively.
The LS-DEM parameters we used for the analyses were 0.6 mm for the surface discretization and 0.25 mm for the volume discretization.
They were chosen based on the convergence study in (\cite{feldfogel_discretization-convergent_2022}).

\subsection{Parametric study and discussion} \label{subsec:Parametric study and discussion}
To isolate the effects of $E$ and $\mu$ on the structural response we varied each of them at a time while keeping the other one constant.
We first consider the effects of $E$ in the range 3.3-75 GPa with a constant $\mu=0.23$ in \ref{subsec:the effects of E}.
Next, we consider the effects of $\mu$ in the range 0.19-9 with a constant $E=4.2$ GPa in \ref{subsec:the effects of mu}.
The constant values of $E$ and $\mu$ are representative ones for polymers and ceramics typically used to manufacture TIS blocks \cite{dyskin_fracture_2003, mirkhalaf_toughness_2019,schaare_point_2008,dalaq_strength_2019,djumas_deformation_2017,djumas_enhanced_2016}.
The explored range of the parameters covers a wide range of materials including concrete (30 GPa) and Aluminum (70 GPa), which have also been used to manufacture TIS blocks \cite{rezaee_javan_impact_2017,schaare_point_2008}.

\subsubsection{The effects of Young's modulus $E$ - scaling} \label{subsec:the effects of E}
\begin{figure}[H] 
    \includegraphics[width=1\textwidth]{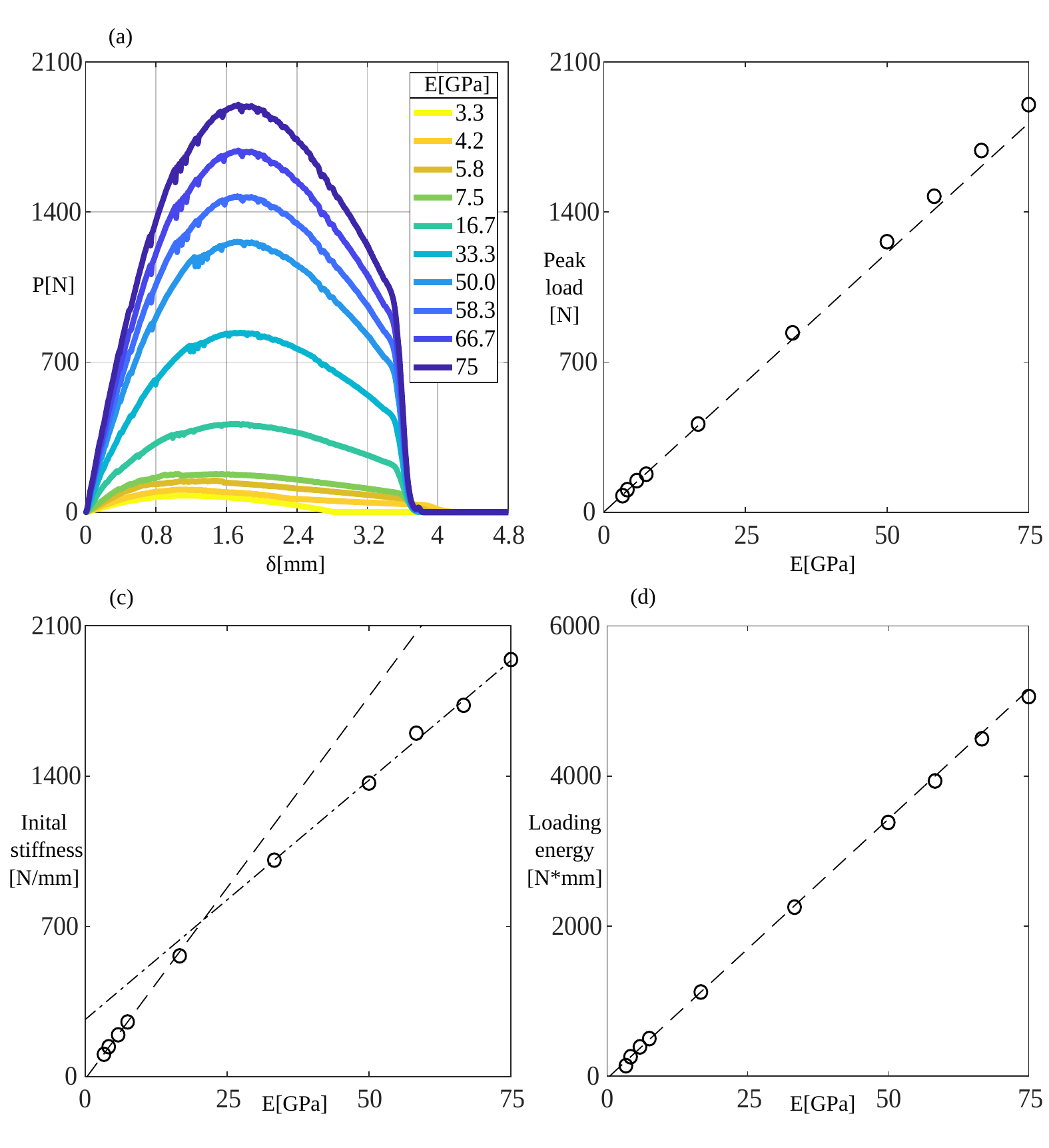} 
    \centering
    \caption{The effects of $E$ on the mechanical behavior and failure of TIS slabs: (a) Load-deflection curves for various $E$'s (in GPa); (b) peak load vs. $E$; (c) global initial stiffness vs. $E$; and (d) loading energy vs. $E$. The dashed and dashed-dotted lines were based on linear interpolation from the five lowest and highest $E$'s, respectively}
    \label{fig:theEffectsOfE_graphs}
\end{figure}

\begin{figure}[H] 
    \includegraphics[width=1\textwidth]{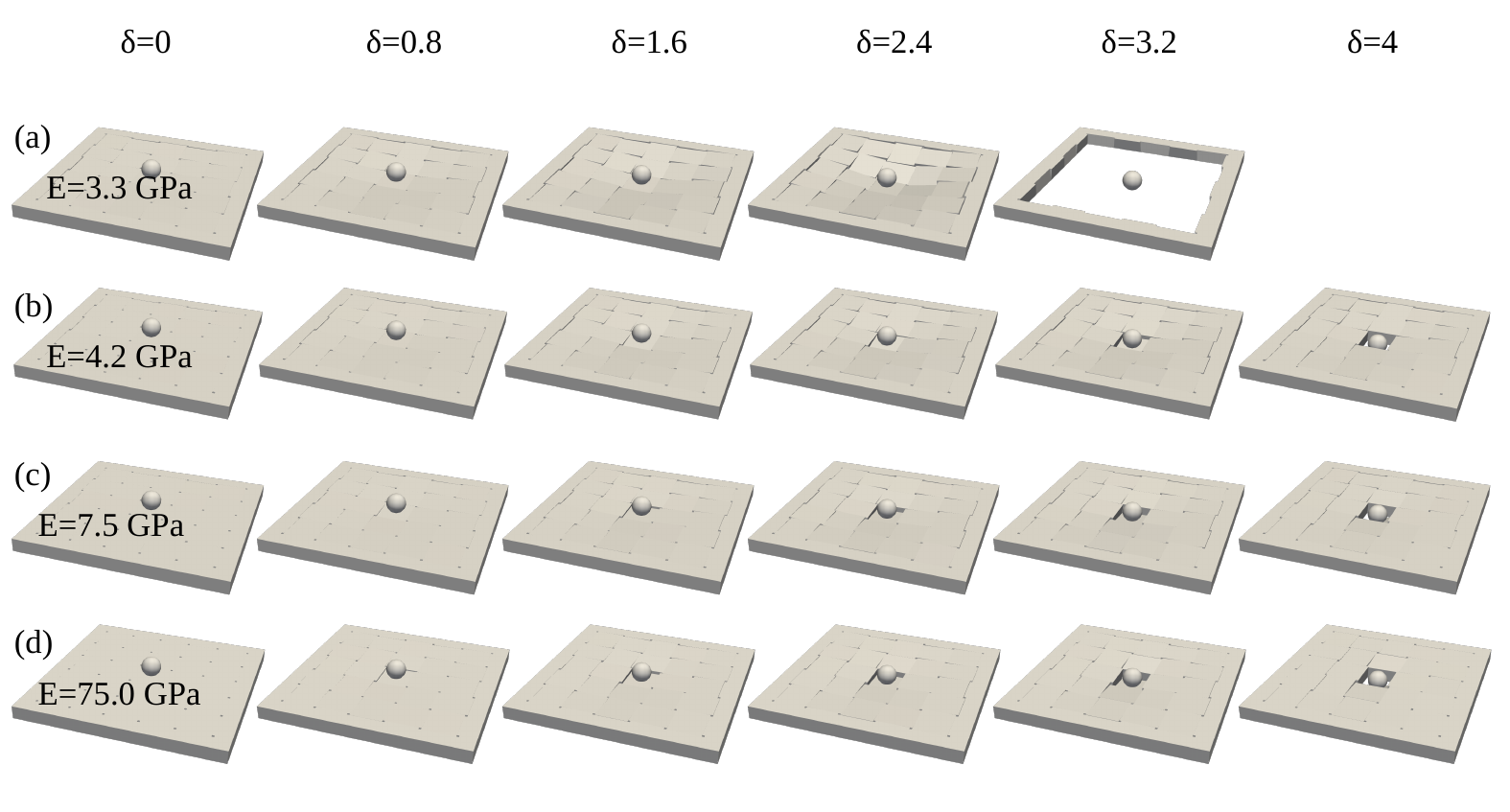} 
    \centering
    \caption{The effects of $E$ on the structural response through sequential snapshots: (a) $E=3.3$ GPa; (b) $E=4.2$ GPa; (c) $E=7.5$ GPa; and (d) $E=75$ GPa.}
    \label{fig:theEffectsOfE_snapshots}
\end{figure}

Fig. \ref{fig:theEffectsOfE_graphs} depicts the effects of $E$ on the $P-\delta$ curves and on the examined response parameters - peak load, initial stiffness, and loading energy.
The structural response is illustrated in Fig. \ref{fig:theEffectsOfE_snapshots} for several $E$'s through snapshots at constant $\delta$ intervals.

The $P-\delta$ curves for $E\geq4.2$ GPa exhibits three main response phases, see Fig. \ref{fig:theEffectsOfE_graphs}a: (I) the load increases linearly up to $\delta\approx0.4$ mm; (II) the stiffness gradually decreases until the peak load is reached at $\delta\approx1.7-1.8$ mm; and (III) the stiffness becomes negative and continues to decrease until an abrupt load drop at $\delta=3.6$ mm brings the load back to zero.
Juxtaposing these phases with the evolution of the structural structural response depicted in Fig. \ref{fig:theEffectsOfE_snapshots}, we identify phase (I) with a mostly stick-governed regime, phase (II) with the initiation and early evolution of slipping mechanisms at the interfaces of the loaded central block, and phase (III) with the response becoming slip-governed and the eventual fall-off of the central block, causing the abrupt load drop at $\delta$=3.6 mm.

The similarlity in failure modes for $E\approx3.7\pm0.5$ GPa suggests that, above a threshhold $E$ (which in our case is somewhere between 3.3 and 4.2 GPa), the failure mode converges to a slip-governed one where only the loaded block ends up falling-off, and where the rest of the structure remains standing.
Furthermore, the higher $E$ is, the less deformed the assembly remains after the central block falls-off.
We note that, although the failure and the regime that precedes it are slip-governed, the peak loads occur either entirely within the stick-governed regime or only slightly after the onset of slip-mechanisms, see the $\delta=1.6$ mm snapshots in Fig. \ref{fig:theEffectsOfE_snapshots}. 

The behavior in the $E$=3.3 GPa case is somewhat different. 
Here, the $P-\delta$ curve is more bell-shaped, it does not exhibit a load drop, and the maximal deflection is notably smaller at $\delta\approx2.7$ mm (Fig. \ref{fig:theEffectsOfE_graphs}a).  
This corresponds to a response that is entirely stick-governed and it leads to a collapse-like failure mode where all the blocks end up falling-off (Fig. \ref{fig:theEffectsOfE_snapshots}a).
The marked difference between the failed configuration for $E\geq4.2$ GPa and the $E$=3.3 GPa case, together with the one-block-fall-off failure for $E\geq4.2$ GPa, speaks to the importance of $E$ in terms of residual carrying capacity.
Here too, the peak loads occurs at $\delta=1.6$ mm, within the stick-governed regime.

\begin{figure}[H] 
    \includegraphics[width=1\textwidth]{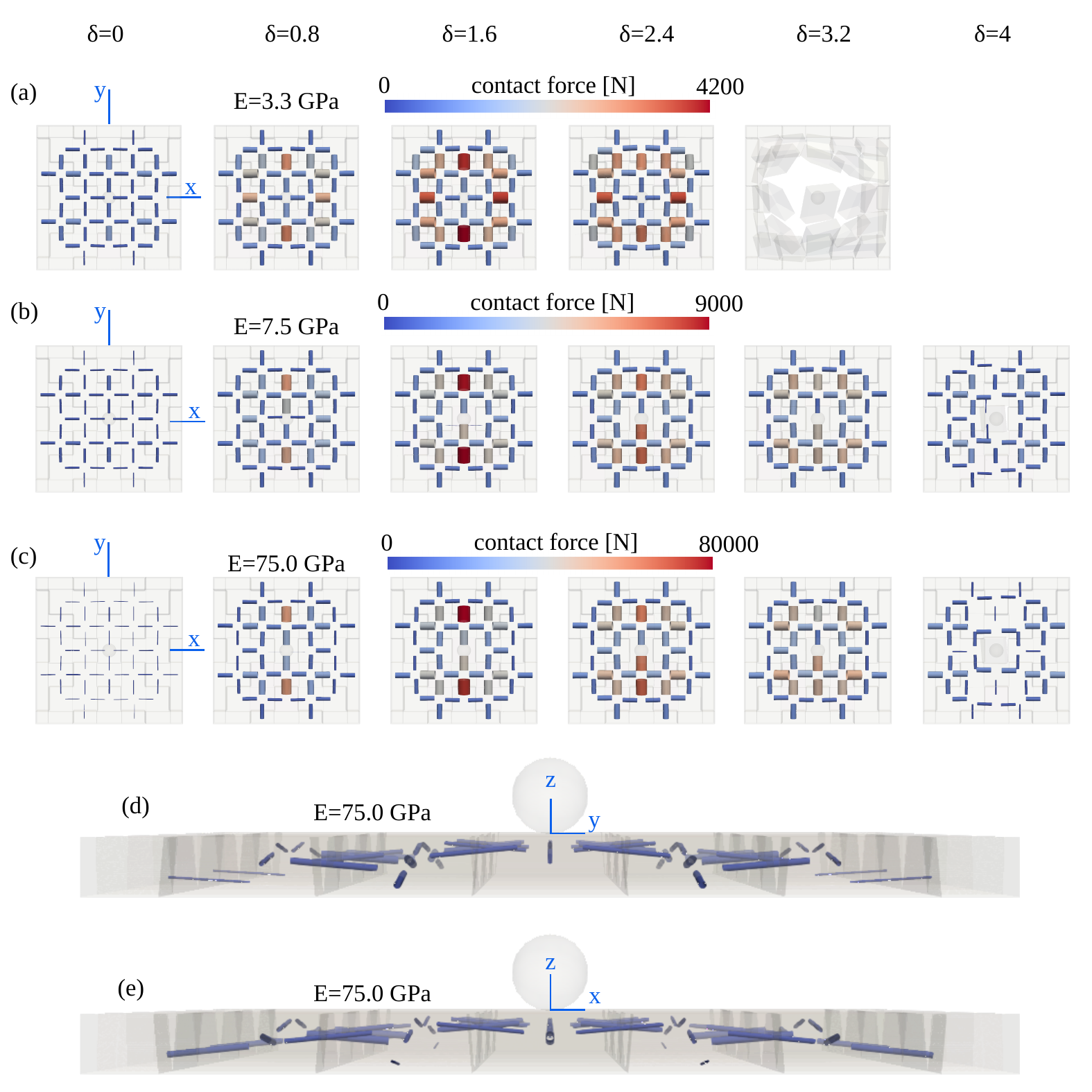} 
    \centering
    \caption{The evolution of contact forces at the interfaces as a function of $E$ through sequential snapshots taken normal to the slab's surface. The contact forces are represented through cylinders whose axes indicates the forces' direction and whose radii indicate their  magnitude: (a) $E=3.3$ GPa; (b) $E=7.5$ GPa; and (c) $E=75$ GPa. Figures (d) and (e) depict the arch-like internal force-chains through two side views for $E=75$ GPa at $\delta$=0 mm.}
    \label{fig:theEffectsOfE_force_chains}
\end{figure}

Regarding the effects of $E$ on the response parameters, both the peak load and the loading energy scale linearly with E, see Fig. \ref{fig:theEffectsOfE_graphs}(b,d).
By comparing Fig. \ref{fig:theEffectsOfE_snapshots}(b-d) to Fig. \ref{fig:theEffectsOfE_snapshots}(a) at $\delta$=1.6 mm, we also observe that the linear scaling of the peak load is unaffected by whether slip-mechanisms at the interfaces of the loaded block have initiated when the peak load is reached or not. 
This can be explained through the idea of thrust-lines that transmit loads through the slab via an internal "truss" whose depth is the effective thickness of the slab, that is, the nominal thickness of the slab minus the deflection under the load. 
Specifically, the transmission of the load to the supports through the interfacial contact forces does not depend on whether there is stick or slip at the interfaces of the loaded block.
Rather, it depends on the magnitude of the deflection of the loaded block, which defines the slab's effective thickness.
This is evidenced by the similarity in the distribution of contact forces at the same prescibed displacement across the range of examined $E$'s, compare Fig. \ref{fig:theEffectsOfE_force_chains}(a-c) at common $\delta$'s. 
We also note the ratio between the maximal contact force and $E$ is more or less the same for the entire range of examined $E$'s.
The arch-like force-chains through which the load is transmitted to the supports are shown from two side views in Fig. \ref{fig:theEffectsOfE_force_chains}(d,e).

The linear scaling of the loading energy is non-trivial and noteworthy because in ordinary structures and materials larger $E$ are typically correlated with decreased energy absorption capabilities.

Unlike the peak load and loading energy, the initial global stiffness scales bilinearly with $E$ (Fig. \ref{fig:theEffectsOfE_graphs}c).
The bilinearity kink occurs at the intersection of the dashed line, which was obtained from linearly interpolating the smallest five $E$'s, and the dahsed-dotted line, which was obtained based on the five highest $E$'s.
We attribute the bilinearity-kink to slipping that occurs for $E\geq20$ GPa at the very first loading stages and that lowered the initial effective thickness, and, commensurately, the global stiffness of the slab.
That this effect indeed was at play is supported by the fact that, per a given $\delta$, more slipping occurs for higher $E$, as can be seen by comparing each of the columns of Fig. \ref{fig:theEffectsOfE_snapshots}.
Specifically, at $\delta$=0.8, mm we see that early slipping occurs for $E$=75 GPa but not for $E=7.5<20$ GPa.
The reason why more slipping occurs for higher $E$'s in the first place is that the ratio $\mu/E$, which represents a normalized resistance to sliding, is inversely related to $E$.

\begin{comment}
Curiously, the ultimate deflection does not scale linearly with $E$ like the other response parameters. 
This can be understood from the shallow-arch/thrust-line analogies for beam-like TIS, wherein when the value of the deflection of the central loaded block reaches the panel's thickness, the structural depth, and the commensurate carrying capacity vanish.
Notably, unlike the 2D case, the ultimate deflection in TI slabs can exceed the panel's thickness, both theoretically and in practice, and this can be seen in many of the examples shown in this manuscript, where the both the experimental and the model's ultimate deflection exceed the h=3.18 mm thickness, e.g., Fig. \ref{fig:validation_50_Pdelta}(i).
This slab's ability to exceed the thickness limit for the ultimate deflection is attributed to alternative load transmission paths that uni-directional beam-like structures lack (namely, transmission in sae to parallel/neighboring strips), but even in TI slabs, the ultimate deflection is more or less dictated by the panel's thickness.
\end{comment}

\subsubsection{The effects of the friction coefficient $\mu$ - saturation} \label{subsec:the effects of mu}
\begin{figure}[H] 
    \includegraphics[width=1\textwidth]{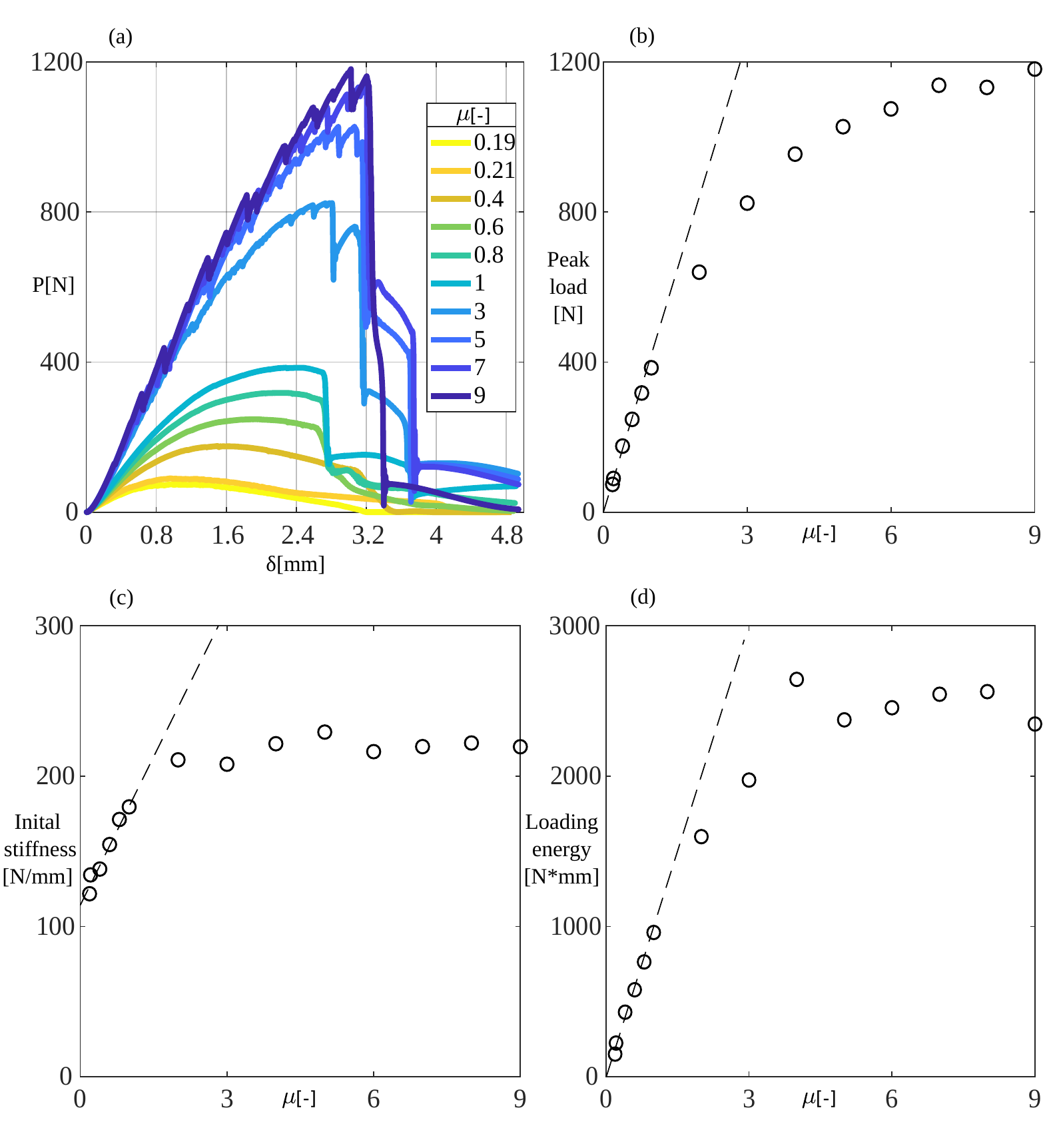} 
    \centering
    \caption{The effects of $\mu$ on the mechanical behavior and failure of TIS slabs: (a) Load-deflection curves for various $\mu$'s; (b) peak load vs. $\mu$; (c) global initial stiffness vs. $\mu$; and (d) loading energy vs. $\mu$. The dashed and dashed-dotted lines were based on linear interpolation from the five lowest and highest $\mu$'s, respectively}
    \label{fig:theEffectsOfMu_graphs}
    \label{fig:the_effects_of_mu_graphs}
\end{figure}

Considering the effects of $\mu$ on the $P-\delta$ curves and the response parameters depicted in Fig. \ref{fig:the_effects_of_mu_graphs}(a), the two most noticeable differences compared with the case of increasing $E$'s are that (1) the pattern of the $P-\delta$ curves changes for larger $\mu$'s, and, most notably (2) the curves eventually converge both in shape and in value for $\mu\approx9$.
In all cases, the stiffness gradually decreases until the peak load is reached, and is followed immediately by an abrupt load drop.
However, whereas the curves for $\mu\leq1$ are bell-like until the first load drop, for $\mu\geq3$ they gradually become more linear, until, as $\mu$ approaches 9 at $\delta\approx3.2$ mm, the steep load drop occurs.
As a results of this change in the shape of the curves, the peak load occurs at larger $\delta$'s for larger $\mu$'s, unlike the case with$E\geq4.2$ GPa, where it is always around $\delta\approx1.7-1.8$ mm.

Regarding the deflection at which the load-drop occurs for $\mu\geq3$, it converges to the thickness of the plate $h=3.18$ mm.
This specific value is explainable in terms of the thrust-line TIS analogy \cite{khandelwal_scaling_2013} by the vanishing of the effective thickness and the zero structural depth at $\delta=h$.
What is interesting here is that, contrary to what the thrust-line model predicts, the slab still maintains a degree of carrying capacity past the load drop at $\delta=h$, as evidenced by the stair-case shapes observed for $\mu\geq1$. We attribute this ability of TIS slabs to sustain loads for $\delta>h$ to the transmission of loads through slip-mechanisms from central rows/columns of blocks to ones closer to the boundary, a feature not shared by one-way-bending TIS beams.

\begin{figure}[H] 
    \includegraphics[width=1\textwidth]{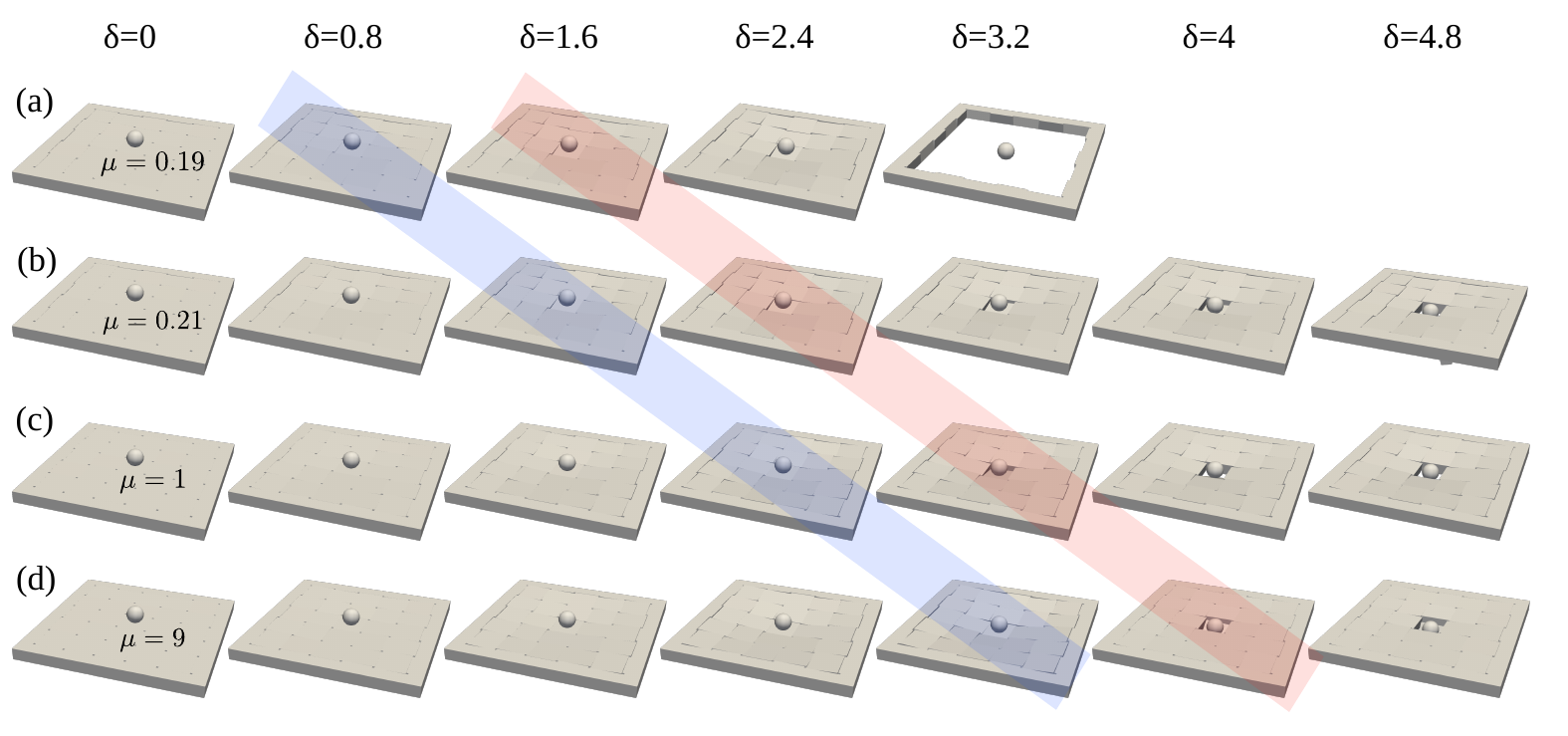} 
    \centering
    \caption{The effects of $\mu$ on the structural response through sequential snapshots: (a) $\mu=0.19$; (b) $\mu=0.21$; (c) $\mu=1$; and (d) $\mu=9$. The diagonal strips indicate the onset of slip-mechanisms and the configuration at which the peak load occurs. The response converges for high $\mu$'s.}
    \label{fig:the_effects_of_mu_snapshots}
\end{figure}

Examining the structural response snapshots in Fig. 
\ref{fig:the_effects_of_mu_snapshots}, we make the following observations:
\begin{enumerate}
    \item Similarly to the effect of $E$, there exist a threshhold $\mu=0.2$ that separates two very different failed configurations. 
    For $\mu\geq0.21$, the response converges in the sense that only the central block falls-off while the rest of the blocks stick and the structures remains standing.
    For $\mu<0.2$, the response is mostly stick-governed and the failure is collapse-like mode where all blocks end up falling-off.
    \item The global deformations that occur during loading are recovered to a degree that depends on how high $\mu$ is. 
    In the limit $\mu=9$, the entire global deformation at $\delta=3.2$ mm is recovered at $\delta=4.8$ mm (Fig. \ref{fig:the_effects_of_mu_snapshots}d). 
    This speaks to $\mu$'s importance to the TIS' residual carrying capacity and global-deformation-recoverability upon un-loading. 
    \item The larger $\mu$ is, the longer the stick-regime is in terms of $\delta$.
    This can be seen in the later onset of slip-mechanism for larger $\mu$'s in the right diagonal strip in Fig. \ref{fig:the_effects_of_mu_snapshots}.
    Recalling that the peak loads occur either entirely within stick-regime or very early after the initiation of slip-mechanisms, the prolonged stick-regime for larger $\mu$'s explains the larger and later-occurring peak load in Fig. \ref{fig:the_effects_of_mu_graphs}(a). 
    This is evidenced by the fact that the $\delta$'s that just precede the onset of slip-mechanisms (shown in the left diagonal strip in Fig. \ref{fig:the_effects_of_mu_snapshots}) are very close to the ones where the peak loads occur for the different $\mu$'s, see Fig. \ref{fig:the_effects_of_mu_graphs}(a).
\end{enumerate}

Turning to $\mu$'s effects on the response parameters, we observe that, following an initial linear increase as a function of $\mu$, all the response parameter plateau or saturation for large $\mu$'s, see Fig. \ref{fig:the_effects_of_mu_graphs}(b-d).
Noting in Fig. \ref{fig:the_effects_of_mu_graphs}(a) that the higher peak load value for $\mu=9$ actually occurs before $\delta=3.2$ mm and within the pre-failure oscillations that occur for the larger $\mu$'s, the peak loads saturates at $\mu\approx7$.
The global stiffness saturates at $\mu=2$ and the loading energy saturates at about $\mu=4-5$.

The small deviations from the saturated loading energy between $\mu=4-9$ are attributed to the post-peak-load "staircases" in the $P-\delta$ curves discussed above in relation to Fig. \ref{fig:the_effects_of_mu_graphs}(a).
These "staircases", which we attribute to the release of load for central rows of blocks and their transmission to adjacent ones closer to the boundary, have a minor effect of the loading energy response parameter and no effect on the other ones.

The saturation of response parameters with $\mu$ has been previously observed in other contexts: saturation of the initial global stiffness in TIS slabs \cite{khandelwal_scaling_2013}, saturation of all response parameters in TIS-like beams \cite{koureas_failure_2022}, and strength saturation in granular systems \cite{karapiperis_investigating_2020,luding_anisotropy_2005}.
Here, we show, for the first time, that not only the stiffness but all aspects of the structural response of TIS slabs saturate as a function of $\mu$.
This includes the peak load, the loading energy, the $P-\delta$ curve, the evolution of the structural response up to and including the failure and post-failure configuration.

In the context of $\mu$ saturation, we note that it occurs at unrealistically high values of $\mu$.
This, however, does not mean that the attainability of a saturated response is similarly unrealistic.
The reason is that the mechanical meaning of $\mu$ saturation is the maximal possible suppression/delaying of slip mechanisms.
This is obtained through maximization of the interface resistance to sliding, of which $\mu$ is but a simplistic and rather crude simplification used for modeling purposes.
The ability to increase the interfacial resistance to sliding through surface architecture, see Djumas \cite{djumas_deformation_2017}, allows to reach the saturated capacities with realistic $\mu$'s \cite{koureas_failure_2022,koureas_beam-like_2022}

\subsubsection{$\mu$ saturation = upper bound on carrying capacity} \label{subsec:upper_bound}

\begin{figure}[H] 
    \includegraphics[width=1\textwidth]{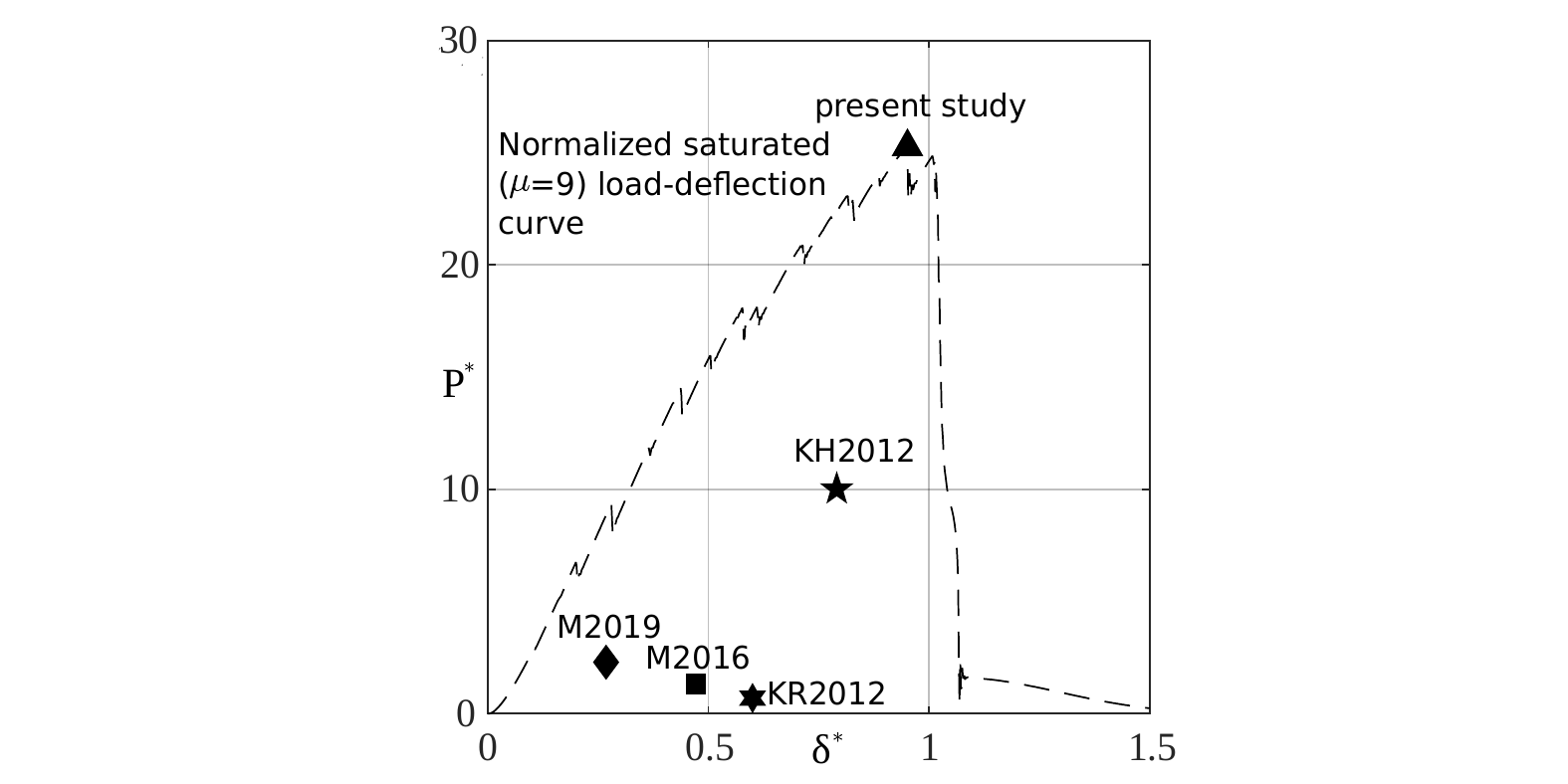} 
    \centering
    \caption{The saturated response is an upper bound on the carrying capacity of TIS slabs: The experimental peak loads in KH2012\cite{khandelwal_transverse_2012}, KR2012\cite{krause_mechanical_2012}, M2016 \cite{mirkhalaf_carving_2016}, and M2019\cite{mirkhalaf_toughness_2019} all fall within the envelope of the normalized saturated load-deflection curve obtained with $\mu=9$. The normalized load $P^{*}$ is taken as $P/E \cdot L^2$ and the normalized deflection $\delta^{*}$ is taken as $\delta/h$ where $E$,$L$, and $h$ are, respectively, the slabs Young's modulus, side length, and thickness.}
    \label{fig:the_saturated_response_is_an_upper_bound}
\end{figure}

To further explore the significance of the saturated structural response shown in Fig. \ref{fig:theEffectsOfE_graphs}, we compared the normalized saturated load-deflection curve and the saturated carrying capacity obtained with $\mu=9$ with all the available experimental results for the carrying capacity of (non-prestressed) centrally loaded TIS slabs we found in the literature \ref{fig:the_saturated_response_is_an_upper_bound}. 
we found that the normalized experimental capacities always fall within the envelope that is the normalized saturated response we found in the present study. 
Since each of the experimental references was taken as the maximal experimental capacity obtained from experimental set-ups reported in that paper, the four reference points actually represent around 60 experimental slabs made with a wide range of materials (glass, ceramics, polymers, Alumina-Silicate), slab lineal dimensions and thicknesses, block shapes (planar and curved-faced), and block types (dense and hollow).
The fact that in such a wide range of experimental benchmarks, the normalized experimental capacities invariably fall under the saturated response envelope suggests that the latter provides an upper bound on the capacity of TIS slabs.
This underscores not only the validity of capacity saturation, but it also speaks to its important physical and structural significance.

\section{Conclusion} \label{sec:Conclusion}
We have presented a systematic study on the various effects of the foremost material properties of Topologically Interlocked Structures (TIS), namely Young's modulus $E$ and the friction coefficient $\mu$, on the structural response of the most common type of TIS application - centrally loaded slabs.

Specifically, we have examined how the load-deflection curve, the structural response parameters - peak load, initial global stiffness, and loading energy, the structural response, the type of failure, and the post-failure configuration depend on $E$ and $\mu$ across a wide and physically representative range of these two parameters.

We have addressed these fundamental, and previously unaddressed, questions using a recently developed modeling approach for TIS \cite{feldfogel_failure_2022} which is based on the Level-Set-Discrete-Element-Method, a uniquely-equipped tool to model the interface-mechanics-governed behavior of TIS slabs.

The main conclusions from our $E-\mu$ parametric study are:
\begin{enumerate}
    \item the peak load and the loading energy scale linearly with $E$, while the initial global stiffness scales bi-linearly with $E$. The maximal deflection converges for large $E$'s.
    \item All response parameters scale initially linearly for small $\mu$'s but later converge, or saturate, for larger $\mu$'s, as does the load-deflection curve.
    \item Above threshold values of $E$ and $\mu$, the failure mechanism converges to a slip-governed one where only the central block falls off and where the residual global deformations become minimal.
    \item Below these thresholds, a collapse-like failure where all blocks end up falling occurs at a relatively small deflection.
    \item For large $\mu$'s, the response converges to one where a major load drop occurs when the deflection equals the thickness of the slab. Following this load drop, the structure still has carrying capacity beyond this deflection and smaller, staircase-like load drops follow. 
    \item The saturated structural response provides an upper bound on the capacity of TIS slabs.
\end{enumerate}

\section{Acknowledgement}
Shai Feldfogel was a Swiss Government Excellence Scholarship holder for the academic years 2021-2022 (ESKAS No. 2021.0165).

\printbibliography

\end{document}